\documentclass[showpacs,preprintnumbers,amsmath,amssymb,superscriptaddress]{revtex4}

\usepackage{amsmath}    
\usepackage{graphicx}   
\usepackage{color}
\usepackage{bm}

\usepackage{dcolumn}
\usepackage{bm}

\usepackage{lineno} 

\begin{document}

\title{$\rho^{0}$ Photoproduction in AuAu Collisions at $\sqrt{s_{NN}}$=62.4~GeV with STAR}

\affiliation{Argonne National Laboratory, Argonne, Illinois 60439, USA}
\affiliation{Brookhaven National Laboratory, Upton, New York 11973, USA}
\affiliation{University of California, Berkeley, California 94720, USA}
\affiliation{University of California, Davis, California 95616, USA}
\affiliation{University of California, Los Angeles, California 90095, USA}
\affiliation{Universidade Estadual de Campinas, Sao Paulo, Brazil}
\affiliation{University of Illinois at Chicago, Chicago, Illinois 60607, USA}
\affiliation{Creighton University, Omaha, Nebraska 68178, USA}
\affiliation{Czech Technical University in Prague, FNSPE, Prague, 115 19, Czech Republic}
\affiliation{Nuclear Physics Institute AS CR, 250 68 \v{R}e\v{z}/Prague, Czech Republic}
\affiliation{University of Frankfurt, Frankfurt, Germany}
\affiliation{Institute of Physics, Bhubaneswar 751005, India}
\affiliation{Indian Institute of Technology, Mumbai, India}
\affiliation{Indiana University, Bloomington, Indiana 47408, USA}
\affiliation{Alikhanov Institute for Theoretical and Experimental Physics, Moscow, Russia}
\affiliation{University of Jammu, Jammu 180001, India}
\affiliation{Joint Institute for Nuclear Research, Dubna, 141 980, Russia}
\affiliation{Kent State University, Kent, Ohio 44242, USA}
\affiliation{University of Kentucky, Lexington, Kentucky, 40506-0055, USA}
\affiliation{Institute of Modern Physics, Lanzhou, China}
\affiliation{Lawrence Berkeley National Laboratory, Berkeley, California 94720, USA}
\affiliation{Massachusetts Institute of Technology, Cambridge, MA 02139-4307, USA}
\affiliation{Max-Planck-Institut f\"ur Physik, Munich, Germany}
\affiliation{Michigan State University, East Lansing, Michigan 48824, USA}
\affiliation{Moscow Engineering Physics Institute, Moscow Russia}
\affiliation{NIKHEF and Utrecht University, Amsterdam, The Netherlands}
\affiliation{Ohio State University, Columbus, Ohio 43210, USA}
\affiliation{Old Dominion University, Norfolk, VA, 23529, USA}
\affiliation{Panjab University, Chandigarh 160014, India}
\affiliation{Pennsylvania State University, University Park, Pennsylvania 16802, USA}
\affiliation{Institute of High Energy Physics, Protvino, Russia}
\affiliation{Purdue University, West Lafayette, Indiana 47907, USA}
\affiliation{Pusan National University, Pusan, Republic of Korea}
\affiliation{University of Rajasthan, Jaipur 302004, India}
\affiliation{Rice University, Houston, Texas 77251, USA}
\affiliation{Universidade de Sao Paulo, Sao Paulo, Brazil}
\affiliation{University of Science \& Technology of China, Hefei 230026, China}
\affiliation{Shandong University, Jinan, Shandong 250100, China}
\affiliation{Shanghai Institute of Applied Physics, Shanghai 201800, China}
\affiliation{SUBATECH, Nantes, France}
\affiliation{Texas A\&M University, College Station, Texas 77843, USA}
\affiliation{University of Texas, Austin, Texas 78712, USA}
\affiliation{University of Houston, Houston, TX, 77204, USA}
\affiliation{Tsinghua University, Beijing 100084, China}
\affiliation{United States Naval Academy, Annapolis, MD 21402, USA}
\affiliation{Valparaiso University, Valparaiso, Indiana 46383, USA}
\affiliation{Variable Energy Cyclotron Centre, Kolkata 700064, India}
\affiliation{Warsaw University of Technology, Warsaw, Poland}
\affiliation{University of Washington, Seattle, Washington 98195, USA}
\affiliation{Wayne State University, Detroit, Michigan 48201, USA}
\affiliation{Institute of Particle Physics, CCNU (HZNU), Wuhan 430079, China}
\affiliation{Yale University, New Haven, Connecticut 06520, USA}
\affiliation{University of Zagreb, Zagreb, HR-10002, Croatia}

\author{G.~Agakishiev}\affiliation{Joint Institute for Nuclear Research, Dubna, 141 980, Russia}
\author{M.~M.~Aggarwal}\affiliation{Panjab University, Chandigarh 160014, India}
\author{Z.~Ahammed}\affiliation{Variable Energy Cyclotron Centre, Kolkata 700064, India}
\author{A.~V.~Alakhverdyants}\affiliation{Joint Institute for Nuclear Research, Dubna, 141 980, Russia}
\author{I.~Alekseev~~}\affiliation{Alikhanov Institute for Theoretical and Experimental Physics, Moscow, Russia}
\author{J.~Alford}\affiliation{Kent State University, Kent, Ohio 44242, USA}
\author{B.~D.~Anderson}\affiliation{Kent State University, Kent, Ohio 44242, USA}
\author{C.~D.~Anson}\affiliation{Ohio State University, Columbus, Ohio 43210, USA}
\author{D.~Arkhipkin}\affiliation{Brookhaven National Laboratory, Upton, New York 11973, USA}
\author{G.~S.~Averichev}\affiliation{Joint Institute for Nuclear Research, Dubna, 141 980, Russia}
% ...

\collaboration{STAR Collaboration}\noaffiliation

\date{\today}

\begin{abstract}

Vector mesons may be photoproduced in relativistic  heavy-ion collisions when a virtual photon emitted by one nucleus scatters from the other nucleus, emerging as a vector meson.   The STAR Collaboration has previously presented measurements of coherent $\rho^0$ photoproduction at center of mass energies of 130 GeV and 200 GeV in AuAu collisions.   Here, we present a measurement of the cross section at 62.4 GeV; we find that the cross section for coherent $\rho^0$ photoproduction with nuclear breakup is $10.5\pm1.5\pm 1.6$ mb at 62.4 GeV.  The cross-section ratio between 200 GeV and 62.4 GeV is $4.4\pm0.6$,  less than is predicted by most theoretical models.  It is, however, proportionally much larger than the previously observed $15$~\% $\pm$ $55$~\% increase between 130 GeV and 200 GeV.

\end{abstract}
\maketitle

\section{\label{sec1}Introduction}

When nuclei cross paths at impact parameters $b$ larger than twice the nuclear radius, $R_A$, they can interact electromagnetically rather than hadronically.
Such events are referred to as Ultra Peripheral Collisions (UPCs). The coupling between 
these  relativistic nuclei is based on
the Weizs\"{a}cker-Williams formalism~\cite{ww}, where their highly boosted electromagnetic fields
are modeled by a flux of photons with small virtuality~\cite{baur}. The photon flux scales
 as the  square of the  nuclear charge and reaches very high values in ions with large atomic number.
   The interaction between the photon flux and nuclear matter is described with an
intermediate fluctuation of the photons into quark antiquark dipoles ($q \overline q$) which then 
scatter from the other nucleus and may emerge as a vector meson. 
The production cross section of the vector meson depends on the $q \overline  q$ coupling to the 
nuclear target.
At small transverse momentum, $p_T < \hbar/R_A$, the  $q \overline q$ pair couples coherently  
to the  entire nucleus. Incoherent coupling takes place at higher $p_T$ where the dipole interacts with 
individual nucleons. The high photon flux allows for several photon exchanges per event. 
Real photons excite giant dipole resonances or higher 
excitation states in the other nucleus which then usually emit one or more neutrons~\cite{baur} in the beam
direction which in turn can be used to trigger on these UPC events.

Four models provide a description of the coherent 
vector meson production in UPC heavy ion interactions. The first one, which we will refer to as KN  
(Klein and Nystrand) is based on vector meson dominance (VMD)  and a classical-mechanical Glauber approach for 
nuclear scattering~\cite{ksjn}. The second model, named FSZ (Frankfurt,  Strikman and  Zhalov),  
makes use of a
 generalized VMD formalism and a QCD Gribov-Glauber approach~\cite{fsz,fsz200}. The third model IIM-GM 
(Iancu, Itakura, Munier - Goncalves, Machado) 
utilizes a QCD color dipole formalism and includes nuclear effects and parton saturation 
phenomena~\cite{gm, iancu}. The fourth  model IPSAT-GM (Impact Parameter Saturation - Goncalves and  Machado) is 
based on the third one but it also  includes the impact parameter dependence of the
dipole interaction with the target nucleus and DGLAP evolution~\cite{dglap} for the target gluon distribution~\cite{gm-sat}. 

The STAR collaboration has previously measured the coherent $\rho^0$ photoproduction in AuAu collisions at center 
of mass per nucleon-pair energies  $\sqrt{s_{NN}}  = 130$ and $200 $  GeV~\cite{meis,STARrho^0}.  STAR observed a rather small energy
 dependence, with only about a $15$~\% $\pm 55$~\% increase in total cross section between the two beam energies.   
 In contrast, both Glauber models (KN and FSZ) predict about a 60\% rise in cross section; 
the two saturation models predict a somewhat slower increase, about 30\% and 33\% respectively.
Here, we present results obtained at a lower energy of $\sqrt{s_{NN}} =  62.4$~GeV to further study the energy-dependence of the cross section. 
We measured the cross section for coherent exclusive $\rho^{0}$ photoproduction accompanied by mutual 
Coulomb excitation of the beam ions and compare the measured cross section with available theoretical 
models.  

\section{\label{sec2}Experimental Setup and Triggering}

The  analysis reported here is based on data collected with the STAR detector from AuAu collisions 
at $\sqrt{s_{NN}}$ = 62.4~GeV  at the Relativistic Heavy Ion Collider at
Brookhaven  National Laboratory.  Charged particles emerging from those interactions have been 
detected with the cylindrical  Time Projection Chamber (TPC)~\cite{tpcdes}. The TPC detected charged  tracks with pseudorapidity $|\eta| < 1.2$
and transverse momentum  $p_T > 100$~MeV/c, with  an 
overall efficiency of  about 85~$\%$. At the time these data  were collected, the TPC was  surrounded 
by the 240 scintillator slats forming the Central Trigger Barrel (CTB)~\cite{trigdes}.   
Two Zero  Degree hadron Calorimeters   (ZDCs)~\cite{zdcdes}  are situated along the beam pipe at $\pm$ 18~m downstream 
 from the interaction  point. The ZDCs  have an  acceptance close to  unity for 
  neutrons  originating from nuclear break-up.

The data were collected with two slightly different minimum bias triggers.  Both triggers required
that the energy in each ZDC be greater than zero, so were sensitive to $\rho^0$ photoproduction
accompanied by mutual Coulomb excitation, while eliminating most cosmic-ray muons, beam-gas
interactions and non beam-beam events.  For the first trigger, trigger~A, this was the only requirement. 
The second trigger, trigger~B, also required that charged particles be detected in the CTB, eliminating empty events, such as those caused by interactions consisting of only mutual Coulomb association.  This trigger had a lower rate, so it was more efficient and could be run with a lower prescale. 

 The data selection criteria applied in this analysis followed closely the ones used in  previous STAR $\rho^0$ photoproduction analyses
 \cite{meis,STARrho^0};  events should have two oppositely charged tracks, each with more than 14  out of 
45 possible hits in the TPC, and both tracks should originate from a common vertex near the
  interaction region. Vertexing efficiency at the level of two tracks 
has been found to be as high as 80~$\%$~\cite{morozov}.  

There are  several types of  backgrounds to $\rho^0$ photoproduction: peripheral
hadronic interactions, other photonuclear interactions, $e^+e^-$ pairs
from two-photon interactions \cite{STARee}, and processes such as beam-gas
interactions, cosmic ray muons and pile-up events. Those events have been suppressed by 
selecting events originating from within a cylindrical region of $15~$cm radius and $100~$cm
longitudinal extension centered at the primary interaction point and containing a $\rho^0$ 
meson candidate with transverse momentum  less than  150 MeV/c. As was the case in previous UPC 
STAR analyses~\cite{meis,STARrho^0}, the contribution from $e^+e^-$ pairs was found to be negligible,  and no particle 
identification was needed  in this analysis due to the low level of  background. 

The geometrical acceptance and the reconstruction efficiency have been studied with the help 
of a Monte Carlo event generator based on the KN model~\cite{STARinterf}. Simulated $\rho^0$ 
photoproduction events have been used as input to the standard STAR detector Monte-Carlo
simulation.  The KN model reproduces the kinematic features of photoproduced $\rho^0$~\cite{STARinterf}, while the detector simulation has been well-tested on central hadronic collisions.  The simulated $\rho^0$ were  then embedded into data collected with a zero-bias trigger (beam bunch crossing-time trigger).  This procedure incorporates the effects of additional tracks, which are present in the data as a result of event pile-up and noise in the STAR TPC. 

The $\rho^{0}$ total reconstruction efficiency has been studied as a function of transverse 
momentum, rapidity, invariant mass  and  azimuthal and polar angles for each of the two trigger implementations. 
Within the rapidity window $|y_{\rho^{0}}|<1$, the mean reconstruction 
efficiency for the data set collected with trigger~A is 36 $\pm$ 3~$\%$ whereas the dataset collected with trigger~B is 9 $\pm$ 1~$\%$. The 
reconstruction efficiencies are relatively constant as functions of transverse momentum, invariant mass, azimuthal and polar angle 
but slowly decrease at higher rapidity ($|y_{\rho^{0}}|>1$), due to the TPC acceptance~\cite{morozov}. 

\section{\label{sec3}$\rho^{0}$ photoproduction}

The sampled luminosity has been determined based on the assumption that the main contribution to 
the total cross section comes from hadronic production with a well known cross section. The luminosity 
was measured by counting the number of events with  more than 313 tracks in the TPC within   
$| y | \leq$  0.5  and with at least ten hits per track. These criteria select 10~$\%$ of the total 
hadronic cross section~\cite{prc_hal}. 
The final measured integrated luminosity for the data selected with trigger~A is 
45 mb$^{-1}$ and for the one accumulated with trigger~B  is 781 mb$^{-1}$. 
The systematic uncertainty for the measured integrated luminosity is 10~$\%$ \cite{STARrho^0}.
Due to the limited luminosity sampled with trigger~A, all results in this publication are based on data 
obtained with trigger~B unless mentioned otherwise.

\begin{figure}[htb]
\centering
\includegraphics[scale=0.4]{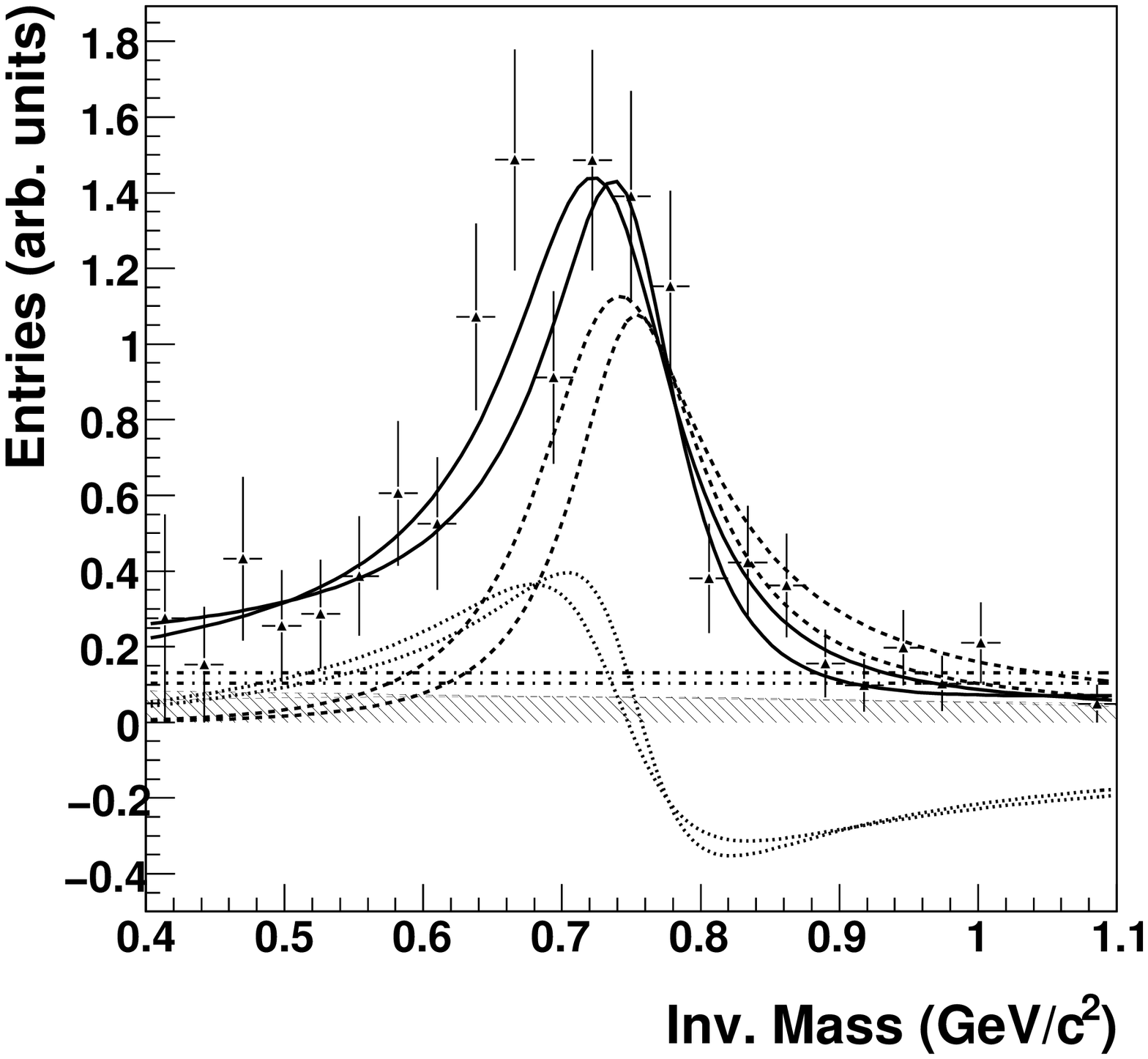}
\includegraphics[scale=0.4]{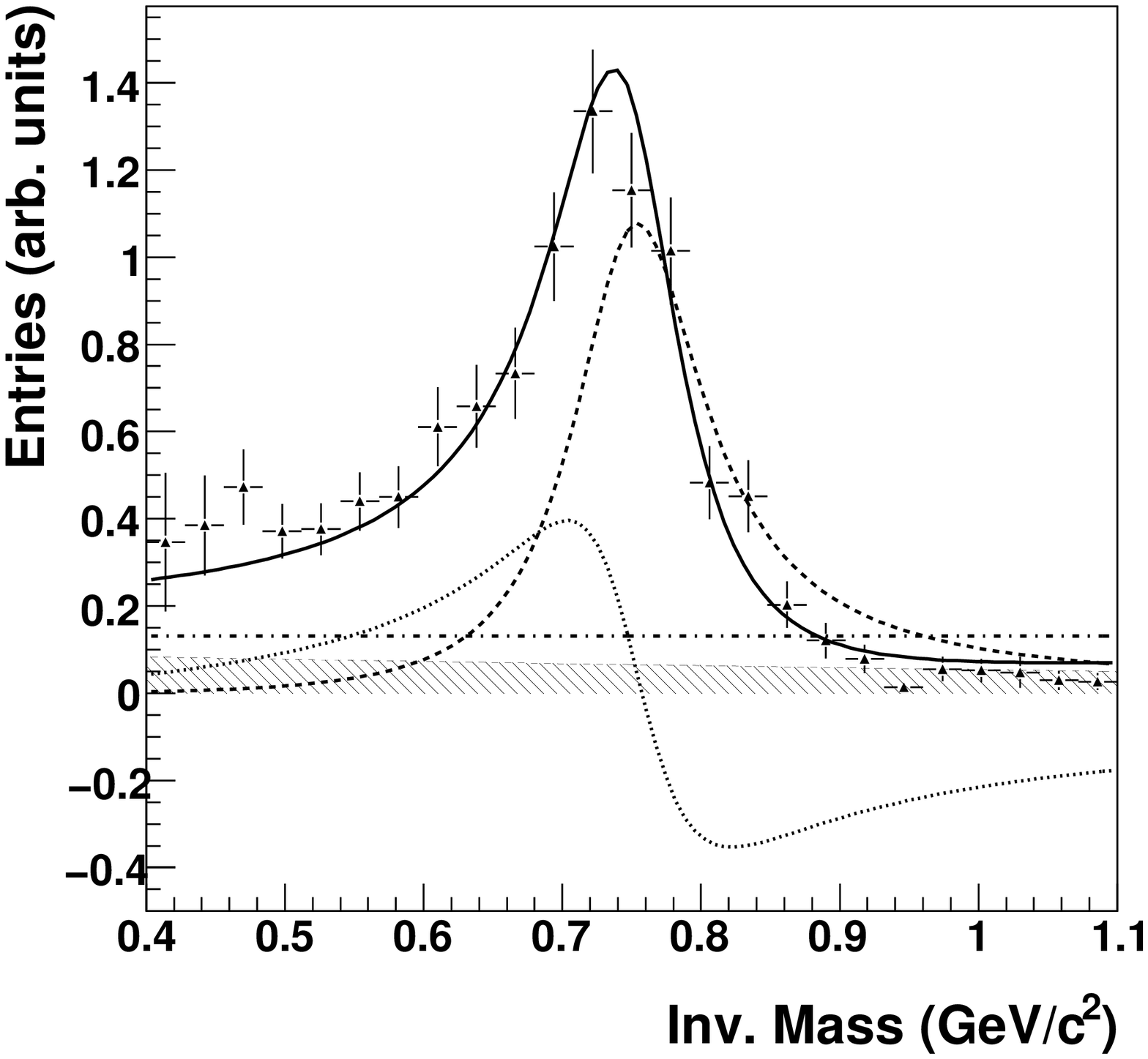}
\caption{\label{massfit}  Acceptance corrected invariant mass distributions for the coherently produced $\rho^{0}$ candidates collected with trigger~A (left) and B (right). The fit function (solid) encompasses   the  Breit-Wigner   (dashed),  the   mass  independent  contribution from direct $\pi^{+}\pi^{-}$ production (dash-dotted),
  and the interference term (dotted). The hatched  area  is   the  contribution  from  the  combinatorial background. The statistical errors are shown.}
\end{figure}

The invariant mass distributions have been obtained with pairs of opposite-sign  charge tracks assumed to
be pions pointing to neutral two-track vertices. The resulting efficiency corrected invariant mass distributions are shown 
in Fig.~\ref{massfit} 
for the data sets collected with trigger~A (left) and trigger~B (right). 

Pion pair photoproduction occurs through two main channels:
pairs are produced by the decay of  
a $\rho^{0}$ meson, or by a photon fluctuating directly into a $\pi^{+}\pi^{-}$ pair,
 which has a flat $M_{\pi^{+}\pi^{-}}$  distribution. These two distributions of pion 
pairs interfere constructively for $M_{\pi^{+}\pi^{-}} < M_{\rho^{0}}$ and destructively for $M_{\pi^{+}\pi^{-}} > M_{\rho^{0}}$.  

\begin{figure}[htb]
\centering
\includegraphics[scale=0.4]{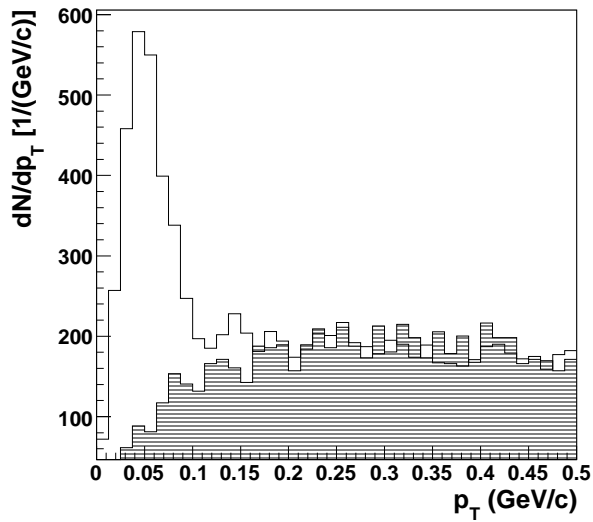}
\caption{\label{rhopt} 
Transverse momentum distribution of the $\rho^{0}$ candidates (open distribution) overlaid by the combinatorial background estimated with like-sign pairs (not corrected to the acceptance and reconstruction efficiency) and scaled to match in the high transverse momentum region, $p_T$ $\ge$ 250 MeV/c (hatched distribution). The plot is based on the dataset collected with trigger~B.}
\end{figure}

Random combinatoric background comes mostly from peripheral AuAu hadronic interactions.
For the coherently produced $\rho^{0}$ mesons, the combinatorial background has been estimated with 
the help of like-sign pairs ($\pi^+\pi^+$ and $\pi^-\pi^-$) and scaled by a factor of 2.2 $\pm$ 0.1 to match the 
unlike pair spectra at high transverse momentum, $p_T$ $\ge$ 250 MeV/c. 
The transverse momentum distribution of $\rho^{0}$ candidates along with the scaled combinatorial background 
are shown in Fig.~\ref{rhopt}.  
  For comparison, we also estimated  the background 
with  high multiplicity events. 
The different methods of background estimation  gave cross sections which differ by less than   3~$\%$.

The invariant mass distribution has been fitted with  a   relativistic  Breit-Wigner function  plus  a
term describing  the   direct  $\pi^{+}\pi^{-}$  production and its interference with the $\rho^{0}$ 
(S\"oding term)~\cite{soding}. 
In addition, a background term was added which was approximated by a second order polynomial. The total fit function then reads:
\begin{equation}\label{eq:fitfunc}
\frac{dN}{dM_{\pi^{+}\pi^{-}}}=\left| A_{\rho^{0}}\frac{\sqrt{M_{\pi^{+}\pi^{-}}M_{\rho^{0}}\Gamma_{\rho^{0}}}}{M_{\pi^{+}\pi^{-}}^{2}-M_{\rho^{0}}^{2}+iM_{\rho^{0}}\Gamma_{\rho^{0}}}+A_{\pi\pi} \right|^{2} + f_{p},
\end{equation}
where the amplitudes for the $\rho^0$ and direct pion pairs are $A_{\rho^{0}}$ and $A_{\pi\pi}$ respectively, and $f_{p}$  is the fixed second order  polynomial which  describes  the background.
The width and mass obtained from the fit to the invariant mass distribution based on the dataset collected with trigger~B are 
 $M_{\rho^{0}}$ = 0.764 $\pm$ 0.009~GeV/c$^2$ and $\Gamma_{\rho^{0}}$ = 0.140 $\pm$ 0.013~GeV/c$^2$ which is consistent with current 
world average (Particle Data Group (PDG))~\cite{pdg} values for photoproduced $\rho^0$ of $M_{\rho^{0}}^{PDG}$ = 0.7685 $\pm$ 0.0011~GeV/c$^2$ and $\Gamma_{\rho^{0}}^{PDG}$ = 0.1507 $\pm$ 0.0029~GeV/c$^2$.

The fit function allows a measurement of the ratio of the  Breit-Wigner 
amplitude ($A_{\rho^{0}}$) to the  amplitude  for  the  direct  $\pi^{+}\pi^{-}$ production ($A_{\pi\pi}$). 
 The measured $|A_{\pi\pi}/A_{\rho^{0}}|$ = 0.88 $\pm$ 0.09 (stat.) $\pm$ 0.09 (syst.)~(GeV/c$^2$)$^{-1/2}$. This is in agreement with previous  STAR measurements of the ratio: 0.81  $\pm$   0.08  (stat.)  $\pm$  0.20 (syst.)~(GeV/c$^2$)$^{-1/2}$ at $\sqrt{s_{NN}}$ = 130~GeV and  0.89 $\pm$ 0.08 (stat.) 
$\pm$ 0.09 (syst.)~(GeV/c$^2$)$^{-1/2}$ at $\sqrt{s_{NN}}$ = 200~GeV.  The systematic errors on the fits 
are determined by varying the fit function 
(for example, by fixing the $\rho^{0}$ mass to the PDG value, or by varying the fitting range and changing the
binning of the invariant mass distribution).

\begin{figure}[htb]
\centering
\includegraphics[scale=0.4]{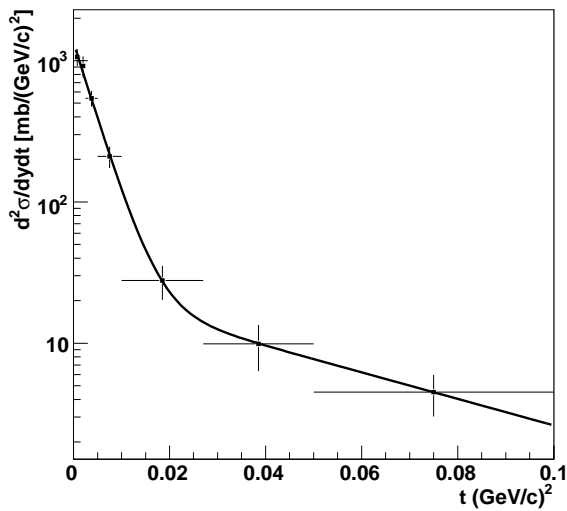}
\caption{\label{dsigmadt} $\rho^0$ production cross section  determined with the data set collected with trigger~B, 
as a function of momentum transfer $t$, fitted with a double exponential fit function in the range $t <$ 0.1. The statistical errors are shown. The ratio of incoherent to coherent cross 
sections is measured to be 0.20 $\pm$ 0.8. The fit parameters are shown in Table~\ref{tableinco}.}
\end{figure}

Figure~\ref{dsigmadt} shows the $\rho^0$ differential production cross section as a function of $t= p_T^2$, averaged over 
rapidity in the $|y_{\rho^{0}}|<1$ window. The  distribution was obtained by fitting invariant mass distributions for each $t$ bin  with Eq. \ref{eq:fitfunc} in order to extract the $\rho^{0}$ yield.  
 The $d^{2}\sigma/dydt$ distribution is fit with the sum of
two exponential terms (see Eq. 5 of Ref.~\cite{STARrho^0}) representing the coherent (small
values of $t$) and the incoherent photoproduction (which dominates for $t > 0.02$ GeV$^{2}$/c$^{2}$). The combinatorial background 
was described by the unscaled like-sign pairs in order to  retain the incoherent $\rho^{0}$ signal. 
 The measured slope for the coherent photoproduction is $B_{\text{coh}}$ =  $257~\pm~32$~(GeV/c)$^{-2}$ 
and the slope for the incoherent photoproduction is $B_{\text{inc}}$ = $21.6 \pm 11.4$ (GeV/c)$^{-2}$.  Both are in agreement with 
 previously published STAR measurements~\cite{STARrho^0}.   The fit was used to determine the ratio  of the  incoherent to coherent 
production cross sections 
$\sigma_{\text{incoherent}}/\sigma_{\text{coherent}}= 0.20 \pm 0.08~\text{(stat.)} \pm 0.08~\text{(syst.)}$ for $|y_{\rho^{0}}|<1$.  This is about $1\sigma$ lower than the measurement $\sigma_{\text{incoherent}}/\sigma_{\text{coherent}}= 0.29 \pm 0.03~\text{(stat.)} \pm 0.08~\text{(syst.)}$ at $\sqrt{s_{NN}}$ = 200~GeV~\cite{STARrho^0}.

\begin{table}
\caption{\label{tableinco}
Parameters for the fit to the $d^{2}\sigma/dydt$. }
\begin{ruledtabular}
\begin{tabular}{ll}
Parameter                         &           \\ \hline  
$A_{\text{coh}}$   &   1328 $\pm$ 159   mb/(GeV/c)$^{2}$  \\
$B_{\text{coh}}$   &   257 $\pm$ 32      (GeV/c)$^{-2}$   \\
$A_{\text{inc}}$    &   22.8 $\pm$ 15.3   mb/(GeV/c)$^{2}$  \\
$B_{\text{inc}}$    &   21.6 $\pm$ 11.4  (GeV/c)$^{-2}$ \\
\end{tabular}
\end{ruledtabular}
\end{table}

The final cross section numbers involve extrapolation from the observed $|y_{\rho^0}|<1$ to all rapidity, and also corrections to account for the presence or absence of accompanying nuclear excitation.  The cross section for coherent production accompanied by mutual nuclear excitation in the $|y_{\rho^0}|<1$ window, measured with data collected with Trigger B, is $\sigma_{\text{coh}} (XnXn, |y_{\rho^{0}}|<1 )$ = 6.2  $\pm$ 0.9 (stat.)  $\pm$ 0.8 (syst.)~mb.  From this, one can infer the cross section in this same rapidity window for coherent production accompanied by the excitation of only a single nucleus, and also for coherent production with no accompanying excitation.   This is done using the ratios  $\sigma(0n0n)/\sigma(XnXn)$ and $\sigma(0nXn)/\sigma(XnXn)$,  where $*n$ represents the number of neutrons 
detected in each ZDC.  These ratios were measured by the STAR collaboration at $\sqrt{s_{NN}}$ = 200~GeV~\cite{STARrho^0} and proved to be in good agreement with those predicted by STARlight. STARlight is a Monte Carlo that simulates two-photon and photon-Pomeron interactions between relativistic nuclei~\cite{STARlight}.  Assuming similar agreement, the ratios predicted by STARlight at $\sqrt{s_{NN}}$ = 62~GeV, $\sigma(0n0n)/\sigma(XnXn)$=  4.6 $\pm$  0.5 (syst.) and $\sigma(0nXn)/\sigma(XnXn)$= 2.7  $\pm$ 0.2 (syst.), are used to extrapolate the current data.  The coherent production cross section with a single nuclear excitation in the $|y_{\rho^0}|<1$ window is then is $\sigma_{\text{coh}}(0nXn,|y_{\rho^{0}}|<1) = 
16.7\ \pm\ 2.7\ \text{(stat.)}\ \pm\ 2.0\ \text{(syst.)}\ \textrm{mb}$, and the coherent production cross section with no accompanying excitation in the $|y_{\rho^0}|<1$ window is $\sigma_{\text{coh}}(0n0n,|y_{\rho^{0}}|<1) = 
28.5\ \pm\ 5.2\ \text{(stat.)}\ \pm\ 4.8\ \text{(syst.)}\ \textrm{mb}$.   

The extrapolation factors to obtain the cross sections over the full rapidity range are model dependent.  Using the KN model~\cite{kn,STARrho^0}, this factor was estimated to be 1.7 $\pm$ 0.1 for events accompanied by nuclear excitation,  1.9 $\pm$ 0.1 for events accompanied by a single nuclear excitation, and 2.7 $\pm$ 0.1 for events with no accompanying excitation.   It is worth noting that, per Fig. 7 of Ref. \cite{STARrho^0}, different models predict rather different extrapolation factors, and the IPOSAT and IIM models would have a considerably smaller extrapolation factor.  

After the extrapolation to full rapidity, we find the total production cross section accompanied by mutual nuclear excitation to be $\sigma_{\text{coh}} (XnXn, \textrm{full-}y) = 10.5\ \pm\  1.5\ \text{(stat.)}\ \pm\ 1.6\ \text{(syst.)}\ \textrm{mb}$, the total production cross section accompanied by a single nuclear excitation to be $\sigma_{\text{coh}}(0nXn, \textrm{full-}y) = 31.8\ \pm\ 5.2\ \text{(stat.)}\ \pm\ 3.9\ \text{(syst.)}\ \textrm{mb}$, and the total production cross section with no accompanying excitation to be $\sigma_{\text{coh}}(0n0n, \textrm{full-}y) = 78\ \pm\ 14 \text{(stat.)}\ \pm\ 13\ \text{(syst.)}\ \textrm{mb}$.  The individual cross sections are summarized in Table~\ref{cross:allnumbers}.

Adding these three cross sections together yields the total coherent cross section at $\sqrt{s_{NN}}$ = 62~GeV, $\sigma_{\text{coh}}(\text{AuAu} \rightarrow \text{Au}^{(*)}\text{Au}^{(*)}\rho^{0})$ = 120 $\pm$  15 (stat.) $\pm$  22 (syst.)$\ \textrm{mb}$. 

We considered several sources for systematic errors as was done in previous work reported in Ref.~\cite{STARrho^0}. 
 The biggest  contributions to the  overall uncertainty in the cross section measurement come from the luminosity measurements, 
the acceptance corrections,  and the extrapolation to the full rapidity range which exhibits strong model dependence. 
The different methods of combinatorial background estimation  gave cross sections which differ by less than   3~$\%$. The contribution due to 
the luminosity measurement is 10~$\%$, and the contribution due to the various cuts is approximately 7~$\%$. 
The error due to the extrapolation to the full phase space is 6~$\%$. Different models used to describe the background 
yielded a 5~$\%$ systematic error.

 The two data sets taken with triggers~A and B were used to cross check measured cross sections and to study systematic effects of different
 trigger requirements on measured luminosity and acceptance corrections. 

The measured cross sections are summarized in Table~\ref{cross:allnumbers} and compared with previous results at  $\sqrt{s_{NN}}$ = 130 and 200~GeV~\cite{meis,STARrho^0}.  The cross-section ratio for the 200 GeV and 62 GeV data is $R=\sigma({\rm 200 GeV})/\sigma({\rm 62.4 GeV}) = 4.4\pm 0.6$, where we have added the systematic and statistical errors in quadrature, neglecting the partial correlation between the sytematic errors at the two energies.     The 130 GeV data have large errors, but we find  $R=\sigma({\rm 130 GeV})/\sigma({\rm 62.4 GeV}) = 3.8\pm 1.9$.    These ratios are much larger than was previously found for $R=\sigma({\rm 200 GeV})/\sigma({\rm 130 GeV}) = 1.15\pm 0.6$, and point to a considerably steeper rise in cross-section with energy, at least as steeply as predicted by the models.    

\begin{table*}
\caption{\label{cross:allnumbers}The total cross section extrapolated to the full rapidity range for coherent $\rho^{0}$ production
  at $\sqrt{s_{NN}}$ =  62.4~GeV accompanied  by nuclear  breakup and  without breakup, compared with  previous measurements at  130~GeV and 200~GeV~\cite{meis,STARrho^0}. The measured cross section, for XnXn events with $|y_{\rho^{0}}|<1$, is based on the dataset collected with trigger~B. Cross sections for other levels of nuclear excitation, and for the full rapidity range, are calculated with extrapolation factors detailed in the text.   Statistical and systematic errors are shown.  }
\begin{ruledtabular}
\begin{tabular}{ccccc}
Parameter                    & STAR at      &  STAR at                     & STAR at             & STAR at                                           \\ 
                             &  $\sqrt{s_{NN}}$ = 62.4~GeV      &  $\sqrt{s_{NN}}$ = 62.4~GeV   & $\sqrt{s_{NN}}$ = 130~GeV \cite{meis} & $\sqrt{s_{NN}}$ = 200~GeV \cite{STARrho^0}       \\
                             &  coherent ($|y_{\rho^{0}}|<1$)   &    coherent (full rapidity)   & coherent (full rapidity)            & coherent  (full rapidity)          \\  \hline
$\sigma^{\rho^{0}}_{XnXn}$ (mb) & 6.2  $\pm$ 0.9  $\pm$ 0.8     & 10.5 $\pm$ 1.5 $\pm$ 1.6   &  28.3 $\pm$ 2.0 $\pm$  6.3   & 31.9 $\pm$ 1.5     $\pm$ 4.5  \\
$\sigma^{\rho^{0}}_{0nXn}$ (mb) & 16.7 $\pm$ 2.7  $\pm$ 2.      &  31.8 $\pm$ 5.2 $\pm$ 3.9  &  95   $\pm$ 60 $\pm$  25     & 105  $\pm$ 5       $\pm$ 15   \\
$\sigma^{\rho^{0}}_{0n0n}$ (mb)  & 28.5 $\pm$ 5.2 $\pm$ 4.8    &  78 $\pm$ 14 $\pm$ 13  &  370  $\pm$ 170 $\pm$ 80     & 391  $\pm$ 18      $\pm$ 55   \\
$\sigma^{\rho^{0}}_{total}$ (mb) & 51.5 $\pm$ 5.9 $\pm$ 5.3    & 120 $\pm$ 15 $\pm$  22  &  460  $\pm$ 220 $\pm$ 110    & 530  $\pm$ 19      $\pm$ 57   \\
\end{tabular}
\end{ruledtabular}
\end{table*}

\begin{figure}[htb]
\centering
\includegraphics[scale=0.8]{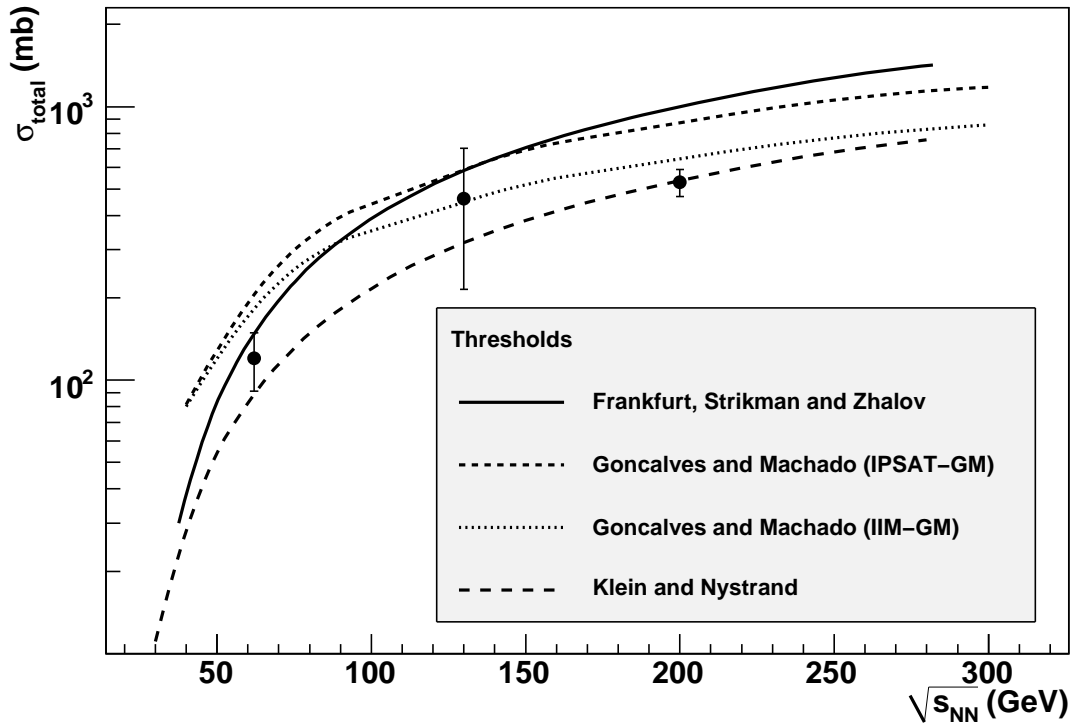}
\caption{\label{crosssection_compare} Comparison of theoretical predictions to the measured total cross section for coherent $\rho^{0}$ 
production as a function of $\sqrt{s_{NN}}$. The measured cross section is based on the dataset collected with trigger~B. The error bars show the sum of the statistical and systematic uncertainties. See text for details. }
\end{figure}

Figure~\ref{crosssection_compare} compares the measured cross section at three different energies with the  aforementioned four theoretical models~\cite{ksjn,fsz,gm}.  For the KN model, the STARlight code was used to predict the energy dependence of the cross-section \cite{STARlight}.  For the other models, we relied on personal communications from the authors to get the energy dependence \cite{pcm,pcbrazil}.  

The rise in cross-section is smaller than is predicted by the KN and FSZ models, which both use Glauber calculations to predict ratios around 6.1  It is closer to the IPSAT-GM and IIM-GM predictions of 3.5 and 4.3 respectively.  However, for these models, the extrapolation from $|y|<1$ to all rapidities would be considerably smaller for the IPSAT and IIM models; since the extrapolation factor depends on the energy, the comparison should be treated with some caution.  For the two Glauber models, the comparison should be more straightforward, although, at 62.4 GeV, the impact parameters are smaller than at 200 GeV, and uncertainties due to the nuclear geometry become more important. 

\section{\label{sec4}Conclusion}

Coherent and incoherent photoproduction of $\rho^{0}$ mesons accompanied by the mutual excitation of the beam ions has been measured in relativistic AuAu collisions at $\sqrt{s_{NN}}$ = 62.4 GeV with the STAR detector at RHIC. The $\rho^{0}$ production 
cross section for the events with mutual excitation ($XnXn$) measured with the 
minimum bias trigger is $10.5\ \pm\  1.5\ \text{(stat.)}\ \pm\ 1.6\ \text{(syst.)}\ \textrm{mb}$ and the total coherent cross section 
$\sigma_{\text{coh}}(\text{AuAu} \rightarrow \text{Au}^{(*)}\text{Au}^{(*)}\rho^{0})$ = 120 $\pm$  15 (stat.) $\pm$  22 (syst.)$\ \textrm{mb}$. The ratio of incoherently to 
coherently produced $\rho^0$ is $0.20 \pm 0.08~\text{(stat.)} \pm 0.08~\text{(syst.)}$, in agreement with previous measurements at higher energies.

The increase in cross section shown by the measurements at 62.4, 130 and 200 GeV is close to the predictions of IIM-GM and somewhat below the predictions of the Glauber theory models.

\begin{acknowledgments}

We thank Mark Strikman, M. Zhalov and M. V. T. Machado for providing theoretical photoproduction cross-sections at 62.4 and 130 GeV. 
We thank the RHIC Operations Group and RCF at BNL, the NERSC Center at LBNL 
and the Open Science Grid consortium for providing resources and support. This 
work was supported in part by the Offices of NP and HEP within the U.S. DOE Office 
of Science, the U.S. NSF, the Sloan Foundation, the DFG cluster of excellence `Origin 
and Structure of the Universe'of Germany, CNRS/IN2P3, FAPESP CNPq of Brazil, Ministry 
of Ed. and Sci. of the Russian Federation, NNSFC, CAS, MoST, and MoE of China, GA and 
MSMT of the Czech Republic, FOM and NWO of the Netherlands, DAE, DST, and CSIR of India, Polish
 Ministry of Sci. and Higher Ed., Korea Research Foundation, Ministry of Sci., Ed. and Sports
 of the Rep. Of Croatia, and RosAtom of Russia.
\end{acknowledgments}

\end{document}